\documentclass[axioms,article,accept,moreauthors,pdftex,12pt,a4paper]{mdpi}
\lastpage{x}
\doinum{10.3390/------}
\pubvolume{xx}
\pubyear{2014}
\history{Received: xx / Accepted: xx / Published: xx}
\Title{Space-time Fractional Reaction-Diffusion Equations Associated with a Generalized Riemann-Liouville Fractional Derivative}

\Author{Ram K. Saxena $^{1}$, Arak M. Mathai $^{2,3}$ and Hans J. Haubold $^{2,4,}$*}

\address{$^{1}$ Department of Mathematics and Statistics, Jai Narain Vyas University, Jodhpur-342005, India;
\linebreak E-Mail: E-Mail: ram.saxena@yahoo.com\\
$^{2}$ Centre for Mathematical and Statistical Sciences, Peechi Campus, KFRI, Peechi-680653, Kerala, India;
\linebreak E-Mail: E-mail: mathai@math.mcgill.ca\\
$^{3}$ Department of Mathematics and Statistics, McGill University, Montreal, Canada;\\
$^{4}$ Office for Outer Space Affairs, United Nations, P.O. Box 500, A-1400 Vienna International Centre, Vienna, Austria}

\corres{E-Mail: hans.haubold@gmail.com; Tel.: +43-676-4252050; Fax: +43-1-26060-5830.}

\abstract{This paper deals with the investigation of the computational solutions of an unified fractional reaction-diffusion equation, which is obtained from the standard diffusion equation by replacing the time derivative of first order by the generalized Riemann-Liouville fractional derivative defined in Hilfer et al. , and the space derivative of second order by the Riesz-Feller fractional derivative, and adding a 
 function $\phi(x,t)$. The solution is derived by the application of the Laplace and Fourier transforms in a compact and closed form in terms of Mittag-Leffler functions. The main result obtained in this paper provides an elegant extension of the fundamental solution for the  space-time fractional diffusion equation obtained earlier by Mainardi et al., and the result very recently given by Tomovski et al.. At the end, extensions of the derived results, associated with a finite number of Riesz-Feller space fractional derivatives, are also investigated.}


\begin{document}

%%%%%%%%%%%%%%%%%%%%%%%%%%%%%%%%%%%%%%%%%%

\section{Introduction}

Standard reaction-diffusion equations are an important class of partial differential equations to investigate nonlinear behavior. Standard nonlinear reaction-diffusion equations can be simulated by numerical techniques.  The reaction-diffusion equation takes into account particle diffusion (different constant and spatial Laplacian operator) and particle reaction (reaction constants and nonlinear reactive terms). Well known special cases of such standard reaction-diffusion equations are the (i) Schloegl model, (ii) Fisher-Kolmogorov equation, (iii) real and complex Ginzburg-Landau equations, (iv) FitzHugh-Nagumo model, and (v) Gray-Scott model. These equations are known under their respective names both in the mathematical and natural sciences literature. The nontrivial behavior of these equations arises from the competition between the reaction kinetics and diffusion. In recent years, interest is developed by several authors in the applications of reaction-diffusion models in pattern formation in physical sciences. In this connection, one can refer to Whilhelmsson and Lazzaro \cite{Wilhelmsson2001}, Hundsdorfer and Verwer \cite{Hundsdorfer2003}, and Sandev et al. \cite{Sandev2011}. These systems show that diffusion can produce the spontaneous formation of spatio-temporal patterns.  For details, see Henry at al. \cite{Henry2002,Henry2005}, and Haubold, Mathai and Saxena \cite{Haubold2007,Haubold2011}.

In this paper, we investigate the solution of an unified model of fractional diffusion system (2.1) in which the two-parameter fractional derivative $D_t^{\mu,\nu}$ sets as a time-derivative and the Riesz-Feller derivative ${_xD_{\theta}^{\alpha}}$ as the space-derivative. This new model provides an extension of the models discussed earlier by \cite{Jespersen1999,DelCastilloNegrete2003,Mainardi2001,Mainardi2005,Kilbas2004,Haubold2007, Saxena2012,Saxena2010,Saxena2006a,Tomovski2012}.

The importance of the derived results further lies in the fact that the Hilfer derivative appeared in the theoretical modeling of broadband dielectric relaxation spectroscopy for glasses, see \cite{Hilfer2002}. For recent and related works on fractional kinetic equations and reaction-diffusion problems, one can refer to papers by \cite{Haubold1995,Haubold2000,Haubold2014,MathaiHaubold2013}, and
\cite{Saxena2002,Saxena2004a,Saxena2004b,Saxena2004c,Saxena2006a,Saxena2006b,Saxena2006c,Saxena2006d,Saxena2008,Saxena2011a,Saxena2011b,Saxena2012,Saxena2012b,Saxena2010}.

%%%%%%%%%%%%%%%%%%%%%%%%%%%%%%%%%%%%%%%%%%

\section{Unified Fractional Reaction-Diffusion Equation}
%% Only for the Journal of Physical and Chemical Gels: Please place the Experimental Section after the Conclusions

In this section, we will derive the solution of the unified reaction-diffusion model
 $$D_t^{\mu,\nu}N(x,t)=\eta~{_xD_{\theta}^{\alpha}}N(x,t)-\omega~N(x,t)+\phi(x,t).\eqno(2.1)
$$
  The main result is given in the form of the following

{\bf Theorem 2.1.}
{\it Consider the unified fractional reaction-diffusion model in (2.1)
were $\eta,t>0,x\in R, \alpha,\theta,\mu,\nu$ are real parameters with the constraints
$$0<\alpha\le 2, |\theta|<\min(\alpha,2-\alpha),\eqno(2.2)
$$and $D_t^{\mu,\nu}$ is the generalized Riemann-Liouville fractional derivative operator defined by (A9) with the conditions
$$I_{0_{+}}^{(1-\nu)(2-\mu)}N(x,0_{+})=f(x),\frac{{\rm d}}{{\rm d}t}I_{0_{+}}^{(1-\nu)(2-\mu)}N(x,0_{+})=g(x),\lim_{|x|\to\infty}N(x,t)=0,\eqno(2.3)
$$where $1<\mu\le 2, 0\le \nu\le 1$, $\omega$ is a constant with the reaction term, ${_xD_{\theta}^{\alpha}}$ is the Riesz-Feller space fractional derivative of order $\alpha$ and symmetry $\theta$ defined by (A11), $\eta$ is a diffusion constant and $\phi(x,t)$ is a  function belonging to the area of reaction-diffusion. Then the solution of (2.1), under the above conditions, is given by
\begin{align}
N(x,t)&=\frac{t^{\mu+\nu(2-\mu)-2}}{2\pi}\int_{-\infty}^{\infty}{\rm e}^{-ikx}f^{*}(k)E_{\mu,\mu+\nu(2-\mu)-1}(-t^{\mu}[\omega+\eta~\psi_{\alpha}^{\theta}(k)]){\rm d}k\nonumber\\
&+\frac{t^{\mu-\nu(\mu-2)-1}}{2\pi}\int_{-\infty}^{\infty}{\rm e}^{-ikx}g^{*}(k)E_{\mu,\mu-\nu(\mu-2)}(-t^{\mu}[\omega+\eta~\psi_{\alpha}^{\theta}(k)]){\rm d}k\nonumber\\
&+\frac{1}{2\pi}\int_{-\infty}^{\infty}\int_0^t\xi^{\mu-1}\phi^{*}(k,t-\xi){\rm e}^{-ikx}\nonumber\\
&\times E_{\mu,\mu}(-\xi^{\mu}[\omega+\eta~\psi_{\alpha}^{\theta}(k)]){\rm d}k{\rm d}\xi.&(2.4)\nonumber
\end{align}}

{\bf Proof:}
If we apply the Laplace transform with respect to the time variable $t$, with Laplace parameter $s$, and Fourier transform with respect to the space variable $x$, with Fourier parameter $k$, use the initial conditions and the formulae (A10) and (A11), then the given equation transforms into the form
\begin{align}
s^{\mu}\tilde{N}^{*}(k,s)&-s^{1-\nu(2-\mu)}f^{*}(k)-s^{\nu(\mu-2)}g^{*}(k)\nonumber\\
&=-\eta~\psi_{\alpha}^{\theta}(k)\tilde{N}^{*}(k,s)-\omega\tilde{N}^{*}(k,s)+\tilde{\phi}^{*}(k,s),\nonumber
\end{align}where according to the convention, the symbol $\tilde{(\cdot)}$ will stand for the Laplace transform of $(\cdot)$ with respect to the time variable $t$ with Laplace parameter $s$ and * will represent the Fourier transform with respect to the space variable $x$ with Fourier parameter $k$. Solving for $\tilde{N}^{*}(k,s)$ we have

$$\tilde{N}^{*}(k,s)=\frac{s^{1-\nu(2-\mu)}f^{*}(k)}{s^{\mu}+\eta~\psi_{\alpha}^{\theta}(k)+\omega}
+\frac{s^{\nu(\mu-2)}g^{*}(k)}{s^{\mu}+\eta~\psi_{\alpha}^{\theta}(k)+\omega}
+\frac{\tilde{\phi}^{*}(k,s)}{s^{\mu}+\eta~\psi_{\alpha}^{\theta}(k)+\omega}.\eqno(2.5)
$$On taking the inverse Laplace transform of (2.5) and applying the result \cite{Saxena2006a} (p. 41)

$$L^{-1}\left[\frac{s^{\beta-1}}{a+s^{\alpha}};t\right]=t^{\alpha-\beta}E_{\alpha,\alpha-\beta+1}(-at^{\alpha}),\eqno(2.6)
$$where $\Re(s)>0,\Re(\alpha)>0,\Re(\alpha-\beta)>-1$, where $\Re(\cdot)$ denotes the real part of $(\cdot)$, it is found that
\begin{align}
N^{*}(k,t)&=f^{*}(k)t^{\mu+\nu(2-\mu)-2}E_{\mu,\mu+\nu(2-\mu)-1}(-t^{\mu}[\omega+\eta~\psi_{\alpha}^{\theta}(k)])\nonumber\\
&+g^{*}(k)t^{\mu+\nu(2-\mu)-1}E_{\mu,\mu+\nu(2-\mu)}(-t^{\mu}[\omega+\eta~\psi_{\alpha}^{\theta}(k)])\nonumber\\
&+\int_0^t\xi^{\mu-1}\phi^{*}(k,t-\xi)E_{\mu,\mu}(-\xi^{\mu}[\omega+\eta~\psi_{\alpha}^{\theta}(k)]){\rm d}\xi.&(2.7)\nonumber\end{align}Taking the inverse Fourier transform of (2.7), we obtain the desired result:
\begin{align}
N(x,t)&=\frac{t^{\mu+\nu(2-\mu)-2}}{2\pi}\int_{-\infty}^{\infty}{\rm e}^{-ikx}f^{*}(k)E_{\mu,\mu+\nu(2-\mu)-1}(-t^{\mu}[\omega+\eta~\psi_{\alpha}^{\theta}(k)]){\rm d}k\nonumber\\
&+\frac{t^{\mu-\nu(\mu-2)-1}}{2\pi}\int_{-\infty}^{\infty}{\rm e}^{-ikx}g^{*}(k)E_{\mu,\mu+\nu(2-\mu)}(-t^{\mu}[\omega+\eta~\psi_{\alpha}^{\theta}(k)]){\rm d}k\nonumber\\
&+\frac{1}{2\pi}\int_{-\infty}^{\infty}\int_0^t\xi^{\mu-1}\phi^{*}(k,t-\xi){\rm e}^{-ikx}\nonumber\\
&\times E_{\mu,\mu}(-\xi^{\mu}[\omega+\eta~\psi_{\alpha}^{\theta}(k)]){\rm d}k{\rm d}\xi.&(2.8)\nonumber\end{align}

%%%%%%%%%%%%%%%%%%%%%%%%%%%%%%%%%%%%%%%%%%

\section{Special Cases of Theorem 2.1.}

(i): If we set $\theta=0$, the Riesz-Feller space derivative reduces to the Riesz space derivative defined by (A15) and consequently, we arrive at the following result:

{\bf Corollary 3.1.}
{\it The solution of the extended fractional reaction-diffusion equation
$$D_t^{\mu,\nu}N(x,t)=\eta~{_xD_0^{\alpha}}N(x,t)-\omega~N(x,t)+\phi(x,t),\eqno(3.1)
$$with  conditions
$$\lim_{|x|\to\infty}N(x,t)=0,0<\alpha\le 2,0\le \nu\le 1,1<\mu\le 2\eqno(3.2)
$$and
$$I_{0_{+}}^{(1-\nu)(2-\mu)}N(x,0_{+})=f(x),\frac{{\rm d}}{{\rm d}t}I_{0_{+}}^{(1-\nu)(2-\mu)}N(x,0_{+})=g(x), x\in R,\eqno(3.3)
$$where $\omega>0$ is a constant with the reaction term, $\eta$ is a diffusion constant, $D_t^{\mu,\nu}$ is the generalized Riemann-Liouville fractional derivative, defined by (A9), ${_xD_0^{\alpha}}$ is the Riesz space fractional derivative operator defined by (A15) and $\phi(x,t)$ is a  function belonging to the area of reaction-diffusion, is given by
\begin{align}
N(x,t)&=\frac{t^{\mu+\nu(2-\mu)-2}}{2\pi}\int_{-infty}^{\infty}{\rm e}^{-ikx}f^{*}(k)E_{\mu,\mu+\nu(2-\mu)-1}(-t^{\mu}[\omega+\eta~|k|^{\alpha}]){\rm d}k\nonumber\\
&+\frac{t^{\mu-\nu(\mu-2)-1}}{2\pi}\int_{-\infty}^{\infty}{\rm e}^{-ikx}g^{*}(k)E_{\mu,\mu+\nu(2-\mu)}(-t^{\mu}[\omega+\eta~|k|^{\alpha}]){\rm d}k\nonumber\\
&+\frac{1}{2\pi}\int_{-\infty}^{\infty}\int_0^t\xi^{\mu-1}\phi^{*}(k,t-\xi)
E_{\mu,\mu}(-\xi^{\mu}[\omega+\eta~|k|^{\alpha}]){\rm e}^{-ikx}{\rm d}k{\rm d}\xi .&(3.4)\nonumber\end{align}}
(ii) For $g(x)=0$, Theorem 2.1 reduces to the following result given by \cite{Saxena2012b}:

{\bf Corollary 3.2.}
{\it Consider the unified fractional reaction-diffusion model
$$D_t^{\mu,\nu}N(x,t)=\eta~{_xD_{\theta}^{\alpha}}N(x,t)-\omega~N(x,t)+\phi(x,t).\eqno(3.5)
$$Here $\eta,t>0,x\in R,\alpha,\theta$ are real parameters with the constraints
$$0<\alpha\le 2, |\theta|<\min(\alpha,2-\alpha).
$$Here  $\omega>0$ is a constant with the reaction term, $D_t^{\mu,\nu}$ is the generalized Riemann-Liouville fractional derivative operator defined by (A9) with the  conditions
$$I_{0_{+}}^{(1-\nu)(2-\mu)}N(x,0_{+})=f(x),~\frac{{\rm d}}{{\rm d}t}I_{0_{+}}^{(1-\nu)(2-\mu)}N(x,0_{+})=0,\lim_{|x|\to\infty}N(x,t)=0,\eqno(3.6)
$$where $x\in R, 1<\mu\le 2, 0\le \nu\le 1$ and ${_xD_{\theta}^{\alpha}}$ is the Riesz-Feller space fractional derivative of order $\alpha$ and symmetry $\theta$ defined by (A11), $\eta$ is a diffusion constant and $\phi(x,t)$is a 
 function belonging to the area of reaction-diffusion. Then the solution of (3.1), under the above conditions, is given by

\begin{align}
N(x,t)&=\frac{t^{\mu+\nu(2-\mu)-2}}{2\pi}\int_{-\infty}^{\infty}{\rm e}^{-ikx}f^{*}(k)E_{\mu,\mu+\nu(2-\mu)-1}(-t^{\mu}[\omega+\eta~\psi_{\alpha}^{\theta}(k)]){\rm d}k\nonumber\\
&+\frac{1}{2\pi}\int_{-\infty}^{\infty}\int_0^t\xi^{\mu-1}\phi^{*}(k,t-\xi){\rm e}^{-ikx}\nonumber\\
&\times E_{\mu,\mu}(-\xi^{\mu}[\omega+\eta~\psi_{\alpha}^{\theta}(k)]){\rm d}k{\rm d}\xi .&(3.7)\nonumber\end{align}}

(iii)  When $\nu=1$, the generalized Riemann-Liouville fractional derivative reduces to Caputo fractional derivative operator, defined by (A6) and we arrive at

{\bf Corollary 3.3.}
{\it Consider the unified fractional reaction-diffusion model
$${^c_0D_t^{\mu}}N(x,t)=\eta~{_xD_{\theta}^{\alpha}}N(x,t)-\omega N(x,t)+\phi(x,t),\eqno(3.8)
$$where $\eta,t>0, x\in R, \alpha,\theta$ are real parameters with the constraints
$$0<\alpha\le 2, |\theta|<\min (\alpha,2-\alpha), 1<\mu\le 2\eqno(3.9)
$$and ${^C_0D_t^{\mu,\nu}}$ is the Caputo fractional derivative operator defined by (A6) with the  conditions
$$N(x,0_{+})=f(x),\frac{{\rm d}}{{\rm d}t}N(x,0_{+})=g(x),\lim_{|x|\to\infty}N(x,t)=0,1<\mu\le 2,\eqno(3.10)
$$where ${_xD_{\theta}^{\alpha}}$ is the Riesz-Feller space fractional derivative of order $\alpha$ and symmetry $\theta$ defined by (A11), $\eta$ is a diffusion constant and $\phi(x,t)$ is a 
 function belonging to the area of reaction-diffusion. Then the solution of (3.8), under the above conditions, is given by
\begin{align}
N(x,t)&=\frac{1}{2\pi}\int_{-\infty}^{\infty}{\rm e}^{-ikx}f^{*}(k)E_{\mu,1}(-t^{\mu}[\omega+\eta~\psi_{\alpha}^{\theta}(k)]){\rm d}k\nonumber\\
&+\frac{1}{2\pi}\int_{-\infty}^{\infty}{\rm e}^{-ikx}g^{*}(k)E_{\mu,2}(-t^{\mu}[\omega+\eta~\psi_{\alpha}^{\theta}(k)]){\rm d}k\nonumber\\
&+\frac{1}{2\pi}\int_{-\infty}^{\infty}\int_0^t\xi^{\mu-1}\phi^{*}(k,t-\xi){\rm e}^{-ikx}\nonumber\\
&\times E_{\mu,\mu}(-\xi^{\mu}[\omega+\eta~\psi_{\alpha}^{\theta}(k)]){\rm d}k{\rm d}\xi.&(3.11)\nonumber\end{align}}

(iv) For $\nu=0$ the fractional derivative $D_t^{\mu,\nu}$ reduces to the Riemann-Liouville fractional derivative operator ${^{RL}_0D_t^{\mu}}$, defined by (A5), and the theorem yields

{\bf Corollary 3.4.}
{\it Consider the unified fractional reaction-diffusion model
$${^{RL}_0D_t^{\mu}}N(x,t)=\eta~{_xD_{\theta}^{\alpha}}N(x,t)-\omega N(x,t)+\phi(x,t),\eqno(3.12)
$$where $\eta,t>0, x\in R,\alpha,\theta$ are real parameters with the constraints
$$0<\alpha\le 2, |\theta|<\min(\alpha,2-\alpha), 1<\mu\le 2,\eqno(3.13)
$$and ${^{RL}_0D_t^{\mu,\nu}}$ is the Riemann-Liouville fractional derivative operator defined by (A5) with the 
 conditions
$${^{RL}_0D_{0_{+}}^{(\mu-2)}}N(x,0_{+})=f(x), {^{RL}_0D_{0_{+}}^{(\mu-1)}}N(x,0_{+})=g(x),\lim_{|x|\to\infty}N(x,t)=0\eqno(3.14)
$$where $x\in R, 1<\mu\le 2$ and ${_xD_{\theta}^{\alpha}}$ is the Riesz-Feller space fractional derivative of order $\alpha$ and symmetry $\theta$ defined by (A11), $\eta$ is a diffusion constant and $\phi(x,t)$ is a 
 function belonging to the area of reaction-diffusion. Then the solution of (3.12), under the above conditions, is given by
\begin{align}
N(x,t)&=\frac{t^{\mu-2}}{2\pi}\int_{-\infty}^{\infty}{\rm e}^{-ikx}f^{*}(k)E_{\mu,\mu-1}(-t^{\mu}[\omega+\eta~\psi_{\alpha}^{\theta}(k)]){\rm d}k\nonumber\\
&+\frac{t^{\mu-1}}{2\pi}\int_{-\infty}^{\infty}{\rm e}^{-ikx}g^{*}(k)E_{\mu,\mu}(-t^{\mu}[\omega+\eta~\psi_{\alpha}^{\theta}(k)]){\rm d}k\nonumber\\
&+\frac{1}{2\pi}\int_{-\infty}^{\infty}\int_0^t\xi^{\mu-1}\phi^{*}(k,t-\xi){\rm e}^{-ikx}\nonumber\\
&\times E_{\mu,\mu}(-\xi^{\mu}[\omega+\eta~\psi_{\alpha}^{\theta}(k)]){\rm d}k{\rm d}\xi.&(3.15)\nonumber\end{align}}

When $\omega\to 0$ then the Theorem 2.1 reduces to the following Corollary which can be stated in the form:

{\bf Corollary 3.5}
{\it Consider the unified fractional reaction-diffusion model
$$D_t^{\mu,\nu}N(x,t)=\eta~{_xD_{\theta}^{\alpha}}N(x,t)+\phi(x,t),\eqno(3.16)
$$where $\eta,t>0, x\in R,\alpha,\theta$ are real parameters with the constraints
$$0<\alpha\le 2, |\theta|<\min(\alpha,2-\alpha), 1<\mu\le 2\eqno(3.17)
$$and $D_t^{\mu,\nu}$ is the generalized Riemann-Liouville fractional derivative operator defined by (A9) with the
 conditions
$$I_{0_{+}}^{(1-\nu)(2-\mu)}N(x,0_{+})=f(x),\frac{{\rm d}}{{\rm d}t}I_{0_{+}}^{(1-\nu)(2-\mu)}N(x,0_{+})=g(x),\lim_{|x|\to\infty}N(x,t)=0,\eqno(3.18)
$$where $x\in R,1<\mu\le 2,0\le \nu\le 1, {_xD_{\theta}^{\alpha}}$ is the Riesz-Feller space fractional derivative of order $\alpha$ and symmetry $\theta$ defined by (A11), $\eta$ is a diffusion constant and $\phi(x,t)$ is a 
 function belonging to the area of reaction-diffusion. Then the solution of (3.16), under the above conditions, is given by
\begin{align}
N(x,t)&=\frac{t^{\mu+\nu(2-\mu)-2}}{2\pi}\int_{-\infty}^{\infty}{\rm e}^{-ikx}f^{*}(k)E_{\mu,\mu+\nu(2-\mu)-1}(-\eta t^{\mu}\psi_{\alpha}^{\theta}(k)){\rm d}k\nonumber\\
&+\frac{t^{\mu+\nu(2-\mu)-1}}{2\pi}\int_{-\infty}^{\infty}{\rm e}^{-ikx}g^{*}(k)E_{\mu,\mu+\nu(2-\mu)}(-\eta t^{\mu}\psi_{\alpha}^{\theta}(k)){\rm d}k\nonumber\\
&+\frac{1}{2\pi}\int_{-\infty}^{\infty}\int_0^t\xi^{\mu-1}\phi^{*}(k,t-\xi){\rm e}^{-ikx}\nonumber\\
&\times E_{\mu,\mu}(-\eta\xi^{\mu}\psi_{\alpha}^{\theta}(k)){\rm d}k{\rm d}\xi.&(3.19)\nonumber\end{align}}

\section{Finite Number of Riesz-Feller Space Fractional Derivatives}

Following similar procedure, we can establish the following:

{\bf Theorem 4.1.}
{\it Consider the unified fractional reaction-diffusion model
$$D_t^{\mu,\nu}N(x,t)=\sum_{j=1}^m\eta_j~{_xD_{\theta_j}^{\alpha_j}}N(x,t)-\omega N(x,t)+\phi(x,t),\eqno(4.1)
$$where $\eta_j,t>0,x\in R,\alpha_j,j=1,...,m,\mu,\nu$ are real parameters, with the constraints
$$1<\mu\le 2, 0\le \nu\le 1,0<\alpha_j\le 2, |\theta_j|\le \min(\alpha_j,2-\alpha_j),j=1,..,m,\eqno(4.2)
$$where $\omega>0$ is coefficient of reaction term, $D_t^{\mu,\nu}$ is the generalized Riemann-Liouville fractional derivative operator defined by (A9) with the 
 conditions

$$I_{0_{+}}^{(1-\nu)(2-\mu)}N(x,0_{+})=f(x), \frac{{\rm d}}{{\rm d}t}I_{0_{+}}^{(1-\nu)(2-\mu)}N(x,0_{+})=g(x),\lim_{|x|\to\infty}N(x,t)=0,\eqno(4.3)
$$where $x\in R,1<\mu\le 2,0\le \nu\le 1, {_xD_{\theta_j}^{\alpha_j}}$ is the Riesz-Feller space fractional derivatives of order $\alpha_j$ and symmetry $\theta_j$ defined by (A11), $\eta_j$ is a diffusion constant and $\phi(x,t)$ is a
 function belonging to the area of reaction-diffusion. Then the solution of (4.1), under the above conditions, is given by
\begin{align}
N(x,t)&=\frac{t^{\mu+\nu(2-\mu)-2}}{2\pi}\int_{-\infty}^{\infty}{\rm e}^{-ikx}f^{*}(k)E_{\mu,\mu+\nu(2-\mu)-1}(-t^{\mu}[\sum_{j=1}^m\eta_j~\psi_{\alpha_j}^{\theta_j}(k)+\omega]){\rm d}k\nonumber\\
&+\frac{t^{\mu-\nu(\mu-2)-1}}{2\pi}\int_{-\infty}^{\infty}{\rm e}^{-ikx}g^{*}(k)E_{\mu,\mu-\nu(\mu-2)}(-t^{\mu}[\sum_{j=1}^m\eta_j~\psi_{\alpha_j}^{\theta_j}(k)+\omega]){\rm d}k\nonumber\\
&+\frac{1}{2\pi}\int_{-\infty}^{\infty}\int_0^t\xi^{\mu-1}\phi^{*}(k,t-\xi){\rm e}^{-ikx}\nonumber\\
&\times E_{\mu,\mu}(-\xi^{\mu}[\sum_{j=1}^m\eta_j~\psi_{\alpha_j}^{\theta_j}(k)+\omega]){\rm d}k{\rm d}\xi.&(4.4)\nonumber\end{align}}

%%%%%%%%%%%%%%%%%%%%%%%%%%%%%%%%%%%%%%%%%%

\section{Special Cases of Theorem 4.1.}

(i)  If we set $\theta_1=\theta_2=...=\theta_m=0$, then by virtue of the identity (A14), we arrive at the following corollary associated with Riesz space fractional derivative:

{\bf Corollary 5.1.}
{\it Consider the unified fractional reaction-diffusion model
$$D_t^{\mu,\nu}N(x,t)=\sum_{j=1}^m\eta_j~{_xD_0^{\alpha_j}}N(x,t)-\omega N(x,t)+\phi(x,t).\eqno(5.1)
$$Here $\eta_j, t>0,x\in R,\alpha_j,\theta_j,j=1,...,m, \mu,\nu$ are real parameters with the constraints
$$0<\alpha_j\le 2,j=1,...,m, 1<\mu\le 2, 0\le \nu\le 1\eqno(5.2)
$$and $D_t^{\mu,\nu}$ is the generalized Riemann-Liouville fractional derivative operator defined by (A9) with the
 conditions
$$I_{0_{+}}^{(1-\nu)(2-\mu)}N(x,0_{+})=f(x), \frac{{\rm d}}{{\rm d}t}I_{0_{+}}^{(1-\nu)(2-\mu)}N(x,0_{+})=g(x),\lim_{|x|\to\infty}N(x,t)=0,\eqno(5.3)
$$where $x\in R, {_xD_0^{\alpha_j}}$ is the Riesz space fractional derivative of order $\alpha_j, j=1,...,m$ defined by (A11), $\omega>0$ is a coefficient of reaction term, $\eta_j$ is a diffusion constant and $\phi(x,t)$ is a 
 function belonging to the area of reaction-diffusion. Then the solution of (5.1), under the above conditions, is given by
\begin{align}
N(x,t)&=\frac{t^{\mu+\nu(2-\mu)-2}}{2\pi}\int_{-\infty}^{\infty}{\rm e}^{-ikx}f^{*}(k)E_{\mu,\mu+\nu(2-\mu)-1}(-t^{\mu}[\sum_{j=1}^m\eta_j~|k|^{\alpha_j}+\omega]){\rm d}k\nonumber\\
&+\frac{t^{\mu-\nu(\mu-2)-1}}{2\pi}\int_{-\infty}^{\infty}{\rm e}^{-ikx}g^{*}(k)E_{\mu,\mu-\nu(\mu-2)}(-t^{\mu}[\sum_{j=1}^m\eta_j~|k|^{\alpha_j}+\omega]){\rm d}k\nonumber\\
&+\frac{1}{2\pi}\int_{-\infty}^{\infty}\int_0^t\xi^{\mu-1}\phi^{*}(k,t-\xi){\rm e}^{-ikx}\nonumber\\
&\times E_{\mu,\mu}(-\xi^{\mu}[\sum_{j=1}^m\eta_j~|k|^{\alpha_j}_+\omega]){\rm d}k{\rm d}\xi.&(5.4)\nonumber\end{align}}

(ii)  Further, if we set $\nu=1$ in the above Theorem 4.1 then the operator $D_t^{\mu,\nu}$ reduces to the Caputo fractional derivative operator ${^C_0D_t^{\mu}}$ defined by (A6), and we arrive at the following result:

{\bf Corollary 5.2.}
{\it Consider the unified fractional reaction-diffusion model
$${^C_0D_t^{\mu}}N(x,t)=\sum_{j=1}^m\eta_j~{_xD_{\theta_j}^{\alpha_j}}N(x,t)-\omega N(x,t)+\phi(x,t),\eqno(5.5)
$$where all the quantities are as defined above with the 
 conditions
$$N(x,0_{+})=f(x), \frac{{\rm d}}{{\rm d}t}N(x,0_{+})=g(x), \lim_{|x|\to\infty}N(x,t)=0\eqno(5.6)
$$and $0<\alpha_j\le 2, 1<\mu\le 2,j=1,...,m, {_xD_{\theta_j}^{\alpha_j}}$ is the Riesz-Feller space fractional derivatives of order $\alpha_j>0, j=1,...,m$, defined by (A11), and $\phi(x,t)$ is a 
 function belonging to the area of reaction-diffusion. Then for the solution of (5.5), there holds the formula
\begin{align}
N(x,t)&=\frac{1}{2\pi}\int_{-\infty}^{\infty}{\rm e}^{-ikx}f^{*}(k)E_{\mu,1}(-t^{\mu}[\sum_{j=1}^m\eta_j~\psi_{\alpha_j}^{\theta_j}(k)+\omega]){\rm d}k\nonumber\\
&+\frac{1}{2\pi}\int_{-\infty}^{\infty}{\rm e}^{-ikx}g^{*}(k)E_{\mu,2}(-t^{\mu}[\sum_{j=1}^m\eta_j~\psi_{\alpha_j}^{\theta_j}(k)+\omega]){\rm d}k\nonumber\\
&+\frac{1}{2\pi}\int_{-\infty}^{\infty}\int_0^t\xi^{\mu-1}\phi^{*}(k,t-\xi){\rm e}^{-ikx}\nonumber\\
&\times E_{\mu,\mu}(-\xi^{\mu}[\sum_{j=1}^m\eta_j~\psi_{\alpha_j}^{\theta_j}(k)+\omega]){\rm d}k{\rm d}\xi.&(5.7)\nonumber\end{align}}
\vskip.1cm\noindent For $m=1, g(x)=0,\omega=0$ the result (5.7) reduces to the one given by \cite{Tomovski2011}.

(iii) If we set $\nu=0$ then the Hilfer fractional derivative defined by (A9) reduces to Riemann-Liouville fractional derivative defined by (A5) and we arrive at the following:

{\bf Corollary 5.3.}
{\it Consider the extended reaction-diffusion model
$${^{RL}_0D_t^{\mu}}N(x,t)=\sum_{j=1}^m\eta_j~{_xD_{\theta_j}^{\alpha_j}}N(x,t)-\omega N(x,t)+\phi(x,t),\eqno(5.8)
$$where the parameters and restrictions are as defined before and with the initial conditions
$${^{RL}_0D_t^{\mu-1}}N(x,0_{+})=f(x),{^{RL}_0D^{(\mu-2)}}N(x,0_{+})=g(x),  \lim_{|x|\to\infty}N(x,t)=0,\eqno(5.9)
$$where $x\in R, 1<\mu\le 2$. Then for the solution of (5.8), there holds the formula
\begin{align}
N(x,t)&=\frac{t^{\mu-2}}{2\pi}\int_{-\infty}^{\infty}{\rm e}^{-ikx}f^{*}(k)E_{\mu,\mu-1}(-t^{\mu}[\sum_{j=1}^m\eta_j~\psi_{\alpha_j}^{\theta_j}(k)+\omega]){\rm d}k\nonumber\\
&+\frac{t^{\mu-1}}{2\pi}\int_{-\infty}^{\infty}{\rm e}^{-ikx}g^{*}(k)E_{\mu,\mu}(-t^{\mu}[\sum_{j=1}^m\eta_j~\psi_{\alpha_j}^{\theta_j}(k)+\omega]){\rm d}k\nonumber\\
&+\frac{1}{2\pi}\int_{-\infty}^{\infty}\int_0^t\xi^{\mu-1}\phi^{*}(k,t-\xi){\rm e}^{-ikx}\nonumber\\
&\times E_{\mu,\mu}(-\xi^{\mu}[\sum_{j=1}^m\eta_j~\psi_{\alpha_j}^{\theta_j}(k)+\omega]){\rm d}k{\rm d}\xi.&(5.10)\nonumber\end{align}}
(iv) Next, if we set $\theta_j=0,j=1,...,m$ in Corollary 5.3 then the Riesz-Feller fractional derivative reduces to Riesz space fractional derivative and we arrive at the following result:

{\bf Corollary 5.4}
{\it Consider the extended fractional reaction-diffusion equation
$${^{RL}_0D_t^{\mu}}N(x,t)=\sum_{j=1}^m\eta_j~{_xD_0^{\alpha_j}}N(x,t)-\omega N(x,t)+\phi(x,t),\eqno(5.11)
$$with the parameters and conditions on them as defined before and with the 
 condition as in (4.1), then for the solution of (5.11) there holds the formula
\begin{align}
N(x,t)&=\frac{t^{\mu-2}}{2\pi}\int_{-\infty}^{\infty}{\rm e}^{-ikx}f^{*}(k)E_{\mu,\mu-1}(-t^{\mu}[\sum_{j=1}^m\eta_j~|k|^{\alpha_j}+\omega]){\rm d}k\nonumber\\
&+\frac{t^{\mu-1}}{2\pi}\int_{-\infty}^{\infty}{\rm e}^{-ikx}g^{*}(k)E_{\mu,\mu}(-t^{\mu}[\sum_{j=1}^m\eta_j~|k|^{\alpha_j}+\omega]){\rm d}k\nonumber\\
&+\frac{1}{2\pi}\int_{-\infty}^{\infty}\int_0^t\xi^{\mu-1}\phi^{*}(k,t-\xi){\rm e}^{-ikx}\nonumber\\
&\times E_{\mu,\mu}(-\xi^{\mu}[\sum_{j=1}^m\eta_j~|k|^{\alpha_j}+\omega]){\rm d}k{\rm d}\xi .&(5.12)\nonumber\end{align}}

(v)  Finally, if we set $g(x)=0,\omega =0$ in Theorem 4.1, it reduces to the one given by \cite{Saxena2012b}. When $m=1$, Corollary 4.1 gives a result given by \cite{Garg2014}.

%%%%%%%%%%%%%%%%%%%%%%%%%%%%%%%%%%%%%%%%%%%% Example of a theorem:

%\begin{Theorem}
%Text text text
%\end{Theorem}

%% Example of a proof:
%\begin{proof}[Proof of Theorem 1]
%Text text text
%\end{proof}

%%%%%%%%%%%%%%%%%%%%%%%%%%%%%%%%%%%%%%%%%%

\section{Conclusions}

In this paper, the authors have presented an extension of the fundamental solution of space-time fractional diffusion given by \cite{Mainardi2001} by using he modified form of the Hilfer derivative given by \cite{Hilfer2009}. The fundamental solution of the equation (2.1) is obtained in closed and computable form. The importance of the results obtained in this paper further lies in the fact that due to the presence of modified Hilfer derivative, results for Riemann-Liouville and Caputo derivatives can be deduced as special cases by taking $\nu=0$ and $\nu=1$ respectively.

%%%%%%%%%%%%%%%%%%%%%%%%%%%%%%%%%%%%%%%%%%

\acknowledgements{Acknowledgments}

The authors would like to thank the Department of Science and Technology, Government of India for the financial support for this work under project No. SR/S4/MS:287/05.

%%%%%%%%%%%%%%%%%%%%%%%%%%%%%%%%%%%%%%%%%%

\authorcontributions{Author Contributions}

All authors contributed to the manuscript. RK Saxena, AM Mathai, and HJ Haubold have contributed to the research methods and the results have been discussed among all authors.

%%%%%%%%%%%%%%%%%%%%%%%%%%%%%%%%%%%%%%%%%%

\conflictofinterests{Conflicts of Interest}

The authors declare no conflict of interest.\par
\medskip\noindent
{\bf Appendix A. Mathematical Preliminaries}

A generalization of the Mittag-Leffler function \cite{MittagLeffler1903,MittagLeffler1905}
$$E_{\alpha}(z)=\sum_{n=0}^{\infty}\frac{z^n}{\Gamma(n\alpha+1)},  \Re(\alpha)>0\eqno(A1)
$$was introduced by \cite{Wiman1905} in the form
$$E_{\alpha,\beta}(z)=\sum_{n=0}^{\infty}\frac{z^n}{\Gamma(n\alpha+\beta)}, \Re(\alpha)>0,\Re(\beta)>0.\eqno(A2)
$$A further generalization of the Mittag-Leffler function is given by \cite{Prabhakar1971} in the following form:
$$E_{\alpha,\beta}^{\gamma}(z)=\sum_{n=0}^{\infty}\frac{(\gamma)_n}{\Gamma(n\alpha +\beta)}, \Re(\alpha)>0,\Re(\beta)>0, \gamma\in C,\eqno(A3)
$$where the Pochhammer symbol is given by
$$(a)_n=a(a+1)...(a+n-1), (a)_0=1,a\ne 0.
$$The main results of the Mittag-Leffler functions defined by (A1) and (A2) are available in the handbook of Erd\'elyi et al. \cite{Erdelyi1955} (Section 18.1) and the monographs of Dzherbashyan \cite{Dzerbashyan1966,Dzerbashyan1993}. The left-sided Riemann-Liouville fractional integral of order $\nu$ is defined by \cite{MillerRoss1993,Samko1990,Kilbas2006,Mathai2010} as
$${^{RL}_0D_t^{-\nu}}N(x,t)=\frac{1}{\Gamma(\nu)}\int_0^t(t-u)^{\nu-1}N(x,t){\rm d}u, t>0, \Re(\nu)>0.\eqno(A4)
$$The left-sided Riemann-Liouville fractional derivative of order $\alpha$ is defined as
$${^{RL}_0D_t^{\alpha}}N(x,t)=(\frac{{\rm d}}{{\rm d}t})^n(I_0^{n-\alpha}N(x,t)), \Re(\alpha)>0, n=[\Re(\alpha)]+1,\eqno(A5)
$$where $[\alpha]$ represents the greatest integer in the real number $x$.
Caputo fractional derivative operator \cite{Caputo1969} is defined in the form
$${^C_0D_t^{\alpha}}f(x,t)=\frac{1}{\Gamma(m-\alpha)}\int_0^t(t-\tau)^{m-\alpha-1}\frac{\partial^m}{\partial t^m}f(x,\tau){\rm d}\tau, m-1<\alpha\le m,\eqno(A6)
$$and it is
$$=\frac{\partial^mf(x,t)}{\partial t^m}, \mbox{ for }\alpha=m, m=1,2,...\eqno(A7)
$$where $\frac{\partial^mf(x,t)}{\partial t^m}$ is the $m$-th derivative of $f(x,t)$ with respect to $t$. When there is no confusion, then the Caputo operator ${^C_0D_t^{\alpha}}$ will be simply denoted by ${_0D_t^{\alpha}}$.
A generalization of the Riemann-Liouville fractional derivative operator (A5) as well as Caputo fractional derivative operator (A6) is given by Hilfer \cite{Hilfer2000} by introducing a left-sided fractional derivative operator of two parameters of order $0<\mu<1$ and type $0\le \nu\le 1$ in the form
$$D_{a^{+}}^{\mu,\nu}N(x,t)=I_{a^{+}}^{\nu(1-\mu)}\frac{\partial}{\partial t}(I_{a^{+}}^{(1-\nu)(1-\mu)}N(x,t)).\eqno(A8)
$$For $\nu=0$, (A8) reduces to the classical Riemann-Liouville fractional derivative operator (A5). On the other hand, for $\nu=1$, it yields the Caputo fractional derivative operator defined by (A6).

{\bf Note A1}:\hskip.3cm The derivative defined by (A8) also occurs in recent papers by \cite{Hilfer2003,Hilfer2009,SrivastavaTomovski2009,Tomovski2009,Tomovski2011,Tomovski2012,Saxena2010,Saxena2012,Garg2014}. Recently, the Hilfer operator defined by (A8) is rewritten in a more general form Hilfer et al. \cite{Hilfer2009} as
$$D_{a^{+}}^{\mu,\nu}N(x,t)=I_{a^{+}}^{\nu(n-\mu)}\frac{\partial^n}{\partial t^n}(I_{0^{+}}^{(1-\nu)(n-\mu)}N(x,t)=I_{a^{+}}^{\nu(n-\mu)}(D_{0^{+}}^{\mu+\nu n-\mu\nu}N(x,t),\eqno(A9)
$$where $n-1<\mu\le n, n\in N, 0\le \nu\le 1$. The Laplace transform of the above operator (A9) is given by Tomovski \cite{Tomovski2011} in the following form:
$$L[D_{a^{+}}^{\mu,\nu}N(x,t);s]=s^{\mu}\tilde{N}(x,s)-\sum_{k=0}^{n-1}s^{s-k-\nu(n-\mu)-1}\frac{\partial^k}{\partial t^k}(I^{(1-\nu)(n-\mu)}N(x,0^{+}),\eqno(A10)
$$for $n-1<\mu\le n, n\in N, 0\le \nu\le 1$. Following Feller \cite{Feller1952}, it is conventional to define the Riesz-Feller space fractional derivative of order $\alpha$ and skewness $\theta$ in terms of its Fourier transform as
$$F\{{_xD_{\theta}^{\alpha}}f(x);k\}=-\psi_{\alpha}^{\theta}(k)f^{*}(k),\eqno(A11)
$$where
$$\psi_{\alpha}^{\theta}(k)=|k|^{\alpha}{\rm e}^{i(sign k)\frac{\theta \pi}{2}}, 0<\alpha\le 2,|\theta|,\min(\alpha,2-\alpha).\eqno(A12)
$$When $\theta=0$, we have a symmetric operator with respect to $x$, that can be interpreted as
$${_xD_0^{\alpha}}=-(-\frac{{\rm d}^2}{{\rm d}x^2})^{\frac{\alpha}{2}}.\eqno(A13)
$$This can be formally deduced by writing $-(k)^{\alpha}=-(k^2)^{\frac{\alpha}{2}}$. For $\theta=0$, we also have
$$F\{{_xD_0^{\alpha}}f(x);k\}=-|k|^{\alpha}f^{*}(k).\eqno(A14)
$$For $0<\alpha\le 2$ and $|\theta|\le \min(\alpha,2-\alpha)$, the Riesz-Feller derivative can be shown to possess the following integral representation in the $x$ domain:
\begin{align}
{_xD_{\theta}^{\alpha}}f(x)&=\frac{\Gamma(1+\alpha)}{\pi}\{\sin[(\alpha+\theta)\frac{\pi}{2}]
\int_0^{\infty}\frac{f(x+\xi)-f(x)}{\xi^{1+\alpha}}{\rm d}\xi\nonumber\\
&+\sin[(\alpha-\theta)\frac{\pi}{2}]\int_0^{\infty}\frac{f(x-\xi)-f(x)}{\xi^{1+\alpha}}{\rm d}\xi\}.\nonumber\end{align}
For $\theta=0$, the Riesz-Feller fractional derivative becomes the Riesz fractional derivative of order $\alpha$ for $1<\lambda\le 2$ defined by analytic continuation in the whole range $0<\alpha\le 2$, $\alpha\ne 1$, see Gorenflo and Mainardi \cite{GorenfloMainardi1999}, as
$${_xD_0^{\alpha}}=-\lambda[I_{+}^{-\alpha}-I_{-}^{-\alpha}],\eqno(A15)
$$where
$$\lambda=\frac{1}{2\cos(\frac{\alpha\pi}{2})}; I_{\pm}^{-\alpha}=\frac{{\rm d}^2}{{\rm d}x^2}I_{\pm}^{2-\alpha}.\eqno(A16)
$$The Weyl fractional integral operators are defined in the monograph by Samko et al. \cite{Samko1990} as
$$(I_{+}^{\beta}N)(x)=\frac{1}{\Gamma(\beta)}\int_{-\infty}^x(x-\xi)^{\beta-1}N(\xi){\rm d}\xi, \Re(\beta)>0
$$and
$$(I_{-}^{\beta}N)(x)=\frac{1}{\Gamma(\beta)}\int_x^{\infty}(\xi-x)^{\beta-1}N(\xi){\rm d}\xi, \Re(\beta)>0.\eqno(A17)
$$
{\bf Note A2}.\hskip.3cm We note that ${_xD_0^{\alpha}}$ is a pseudo differential operator. In particular, we have
$${_xD_0^2}=\frac{{\rm d}^2}{{\rm d}x^2},\mbox{  but  } {_xD_0^1}\ne \frac{{\rm d}}{{\rm d}x}.\eqno(A18)
$$

%%%%%%%%%%%%%%%%%%%%%%%%%%%%%%%%%%%%%%%%%%

%=================================================================
% References: Variant A
%=================================================================
% Back Matter (References and Notes)
%----------------------------------------------------------
% Style and layout of the references
\bibliographystyle{mdpi}
% \makeatletter
% \renewcommand\@biblabel[1]{#1. }
% \makeatother

%=================================================================
% References:  Variant B
%=================================================================
% Use the following option to include external BibTeX files:
%\bibliography{lite}
%\bibliographystyle{mdpi}

\end{document}